\documentstyle[11pt,aaspp]{article}

\def\et{{\it et al. }}
\tighten

\articleid{11}{14}
\slugcomment{Submitted to Ap.J.}
\begin{document}

\title{ $\gamma$-Ray Pulsars: Emission Zones and Viewing Geometries}

\author{Roger W. Romani\altaffilmark{1} and I.-A. Yadigaroglu}
\affil{Department of Physics, Stanford University,
    Stanford, CA 94305-4060}

\altaffiltext{1}{Alfred P. Sloan Fellow}

\begin{abstract}
	There are now a half dozen young pulsars detected in high energy
photons by the Compton GRO, showing a variety of emission efficiencies
and pulse profiles.  We present here a
calculation of the pattern of high energy emission on the sky in a
model which posits $\gamma$-ray production by charge depleted gaps
in the outer magnetosphere. This model accounts for the
radio to $\gamma$-ray pulse offsets of the known pulsars, as well as the
shape of the high energy pulse profiles. We also show that $\sim 1/3$
of emitting young radio pulsars will not be detected due to beaming
effects, while $\sim 2.5 \times$ the number of radio-selected $\gamma$-ray
pulsars will be viewed only high energies.  Finally we compute the
polarization angle
variation and find that the previously misunderstood optical
polarization sweep of the Crab pulsar arises naturally in this picture.
These results strongly support an outer-magnetosphere location for the
$\gamma-$ray emission.
\end{abstract}


\section{Introduction}

	Since the birth of $\gamma$-ray astronomy the two most prominent
galactic point sources have been identified with young rotation-powered
pulsars, the Crab and Vela pulsars. The site of $\gamma$-ray production
has however been intensely debated for almost two decades. The two
principal models are the `polar cap' picture exemplified by the work
of Daugherty and Harding (1982) and the `outer gap' model championed
by Cheng, Ho and Ruderman (1988, CHR). In the former acceleration occurs near
the neutron star surface and the observed emission results from a
$\gamma-B$ pair cascade. In the second model acceleration is posited in
charge depleted regions near the speed of light cylinder and
$\gamma-\gamma$ pair production is an important process. We have
examined these two models (Chiang and Romani 1992) and find that the
observed pulse profiles arise most naturally in a modified version of
the outer gap picture. Moreover, spectral calculations (Chiang and
Romani 1994, CR94) have shown that in this model emission processes vary
throughout the magnetosphere and can
produce spectral variations through the pulse like those seen. In this
paper we quantitatively compare the outer gap predictions with the
observed profiles, supporting these conclusions.

	The impetus for these computations is the dramatic increase in
information on the $\gamma$-ray pulsars provided by the {\it Compton
Gamma Ray Observatory}. In addition to improved light curves and phase
resolved spectra for Crab and Vela ({\it e.g.} Nolan \et  1993),
CGRO has detected at least four
other pulsars in high-energy emission, including Geminga, which has
not been seen in radio emission ({\it e.g.} Thompson, \et 1992,
Ulmer, \et 1993, Fierro, \et 1993,
Mayer-Hasselwander \et 1994). Other bright plane sources await
confirmation as pulsars, and the expected number of each of these
classes of objects is presently unknown. It is apparent from these data that
older rotation-powered pulsars are increasingly efficient producers of
$\gamma$ emission, at least for characteristic ages $\la 10^{5.5} -
10^6$y. In deciphering the puzzle of $\gamma$ production efficiencies
upper limits on other pulsars are also important. Their interpretation
is, however, unclear. Finally detailed information on the location of
the emission region is an essential tool for understanding the
polarization and
spectrum seen from the observed pulsars. The geometrical computations
of pulsar beaming in this paper address each of these issues.

\section{Emission Zone Geometry and Profile Calculation}

	Well above the neutron star surface the dipole component of the
magnetic field will dominate. We have computed the location of the
outer magnetosphere charge depletion regions (outer gaps) by calculating
the dipole field structure from the full retarded potentials. The
outer gaps (CHR) are regions above the null charge surface (where the
co-rotation charge density changes sign) extending to the speed of light
cylinder, where co-rotation must break down. Accordingly plasma lost
from this gap cannot be replenished from the stellar surface and
a charge depleted region with significant ${\bf E \cdot B}$ is developed.
This is the location of the particle acceleration which gives rise
to the observed $\gamma$-ray emission. Following CHR, we assume that
the outer gap will be bounded below by the surface of last closed field lines.
We determine this surface by finding the dipole field lines which
are tangent to speed of light cylinder. Traced back to the stellar surface,
these give a polar cap moderately extended in longitude for significant
magnetic inclinations $\alpha$.
The upper gap surface will be along field lines for which substantial
$\gamma-\gamma$ pair production shorts out the outer gap. It is this
upper surface, where the population of accelerated pairs builds up to
appreciable densities, that in fact constitutes the emission zone of the
gap.

	This emission surface lies significantly within the open field
line region above the polar cap. We find ($c.f.$ Chiang and Romani 1994)
the soft photons supporting the $\gamma-\gamma$ pair production that
limits gap growth are provided principally from regions in the inner
magnetosphere (the lowest portions of the gap, where synchrotron losses
are significant) or from the stellar surface. Pair production will
be strongest close to this soft photon source, and so we assume that
the gap width for a given position around the polar cap is determined by
this pair production at the null charge surface. Two effects determine the
pair production length and hence the gap width at this point. First as
one moves to the sides of the polar cap, the intersection of the null charge
surface with the last closed line moves higher in the magnetosphere.
There the soft photon flux at the threshold for pair creation on the
curvature $\gamma$ from primaries accelerated to radiation reaction limited
energies in the gap will decrease -- this tends to increase the pair
production length.  Countering this to some extent, the angle between
the central soft flux and the locally produced high energy $\gamma$'s will
increase higher in the magnetosphere -- this lowers the pair creation threshold
and for a steep soft spectrum from thermal or synchrotron radiation,
decreases the pair production length. The complete solution of the
gap growth and radiation production is a complex non-linear problem
(CR94). Here we adopt a simple scaling law for
the gap width $w$ at the null charge surface which approximates these
\begin{figure}
\plotfiddle{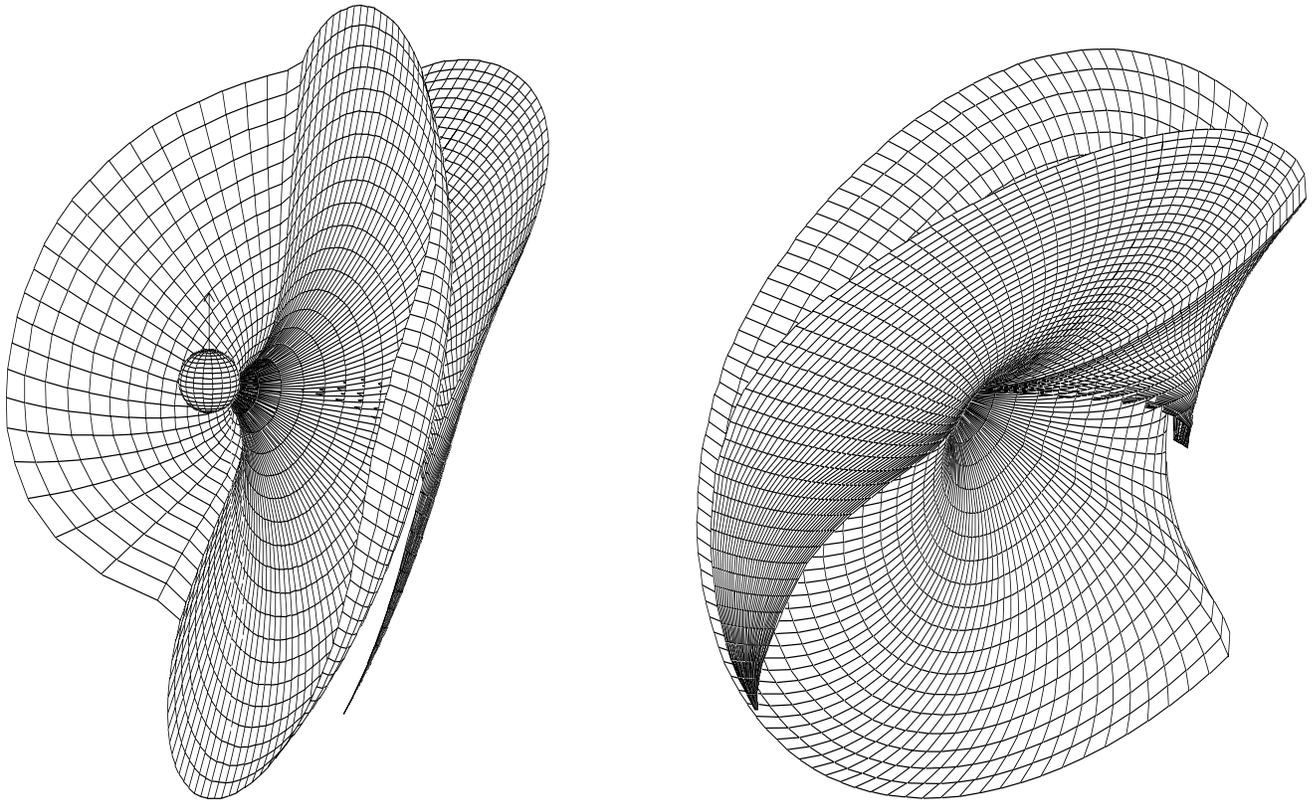}{3.9truein}{-90}{70}{70}{-260}{315}
\caption{ Two views of the surface of last closed field lines, cut at $1.2
r_{LC}$ for an $\alpha=70^\circ$ rotator (see text).
The gap lies above the null charge surface, inside the open field line
region. We show the illuminated section of the the gap upper surface.
The observed pulsar $\gamma$-rays are beamed along this surface. The neutron
star's radius is increased to $0.1r_{LC}$ for clarity.
}
\label{gapsurf}
\end{figure}
behaviors $w \propto r_c$, with $r_c$ the local field line radius of curvature
and with the width at the gap's lowest point scaling as $\alpha^{-1}$.
This lowest width can also depend on the $P$ and $B$, thereby controlling
the pulsar luminosity -- this has not been treated in detail.
Thus gap regions starting at high altitude will be wider. However, the
gap width does not grow as fast as the $B \sim r^{-3}$ spreading of the field
lines; the flux threaded towards the sides of the gap decreases. Along any
given field line, charges will flow above the null charge surface, so
we take the set of field lines determined above to define the gap upper
surface (Figure~\ref{gapsurf}).

	We note that we have not included the effects of particle inertia on
the field lines or the field perturbations induced by magnetospheric currents.
These should be small in our near vacuum region as long as we are not very
close to the light cylinder. In particular these produce only small changes in
the geometry of the polar cap. Moreover, we wish to emphasize that the majority
of our results are quite robust to perturbations of the field geometry
and are thus independent of the details of how we have chosen our gap
surface. However, fine structure in observed pulse profiles can help detail
the precise location of the high energy emission zone. The brightness
along the gap builds up as the particle density increases, but then falls off
as the light cylinder is approached. As an estimate of this variation we take
$ F \propto 2s - s^2 $, where $s$ is the arc length along the field line
measured in units of the light cylinder radius, and impose an exponential
cutoff at $s=1$ with $\sigma=0.5$.

\begin{figure}
\plotfiddle{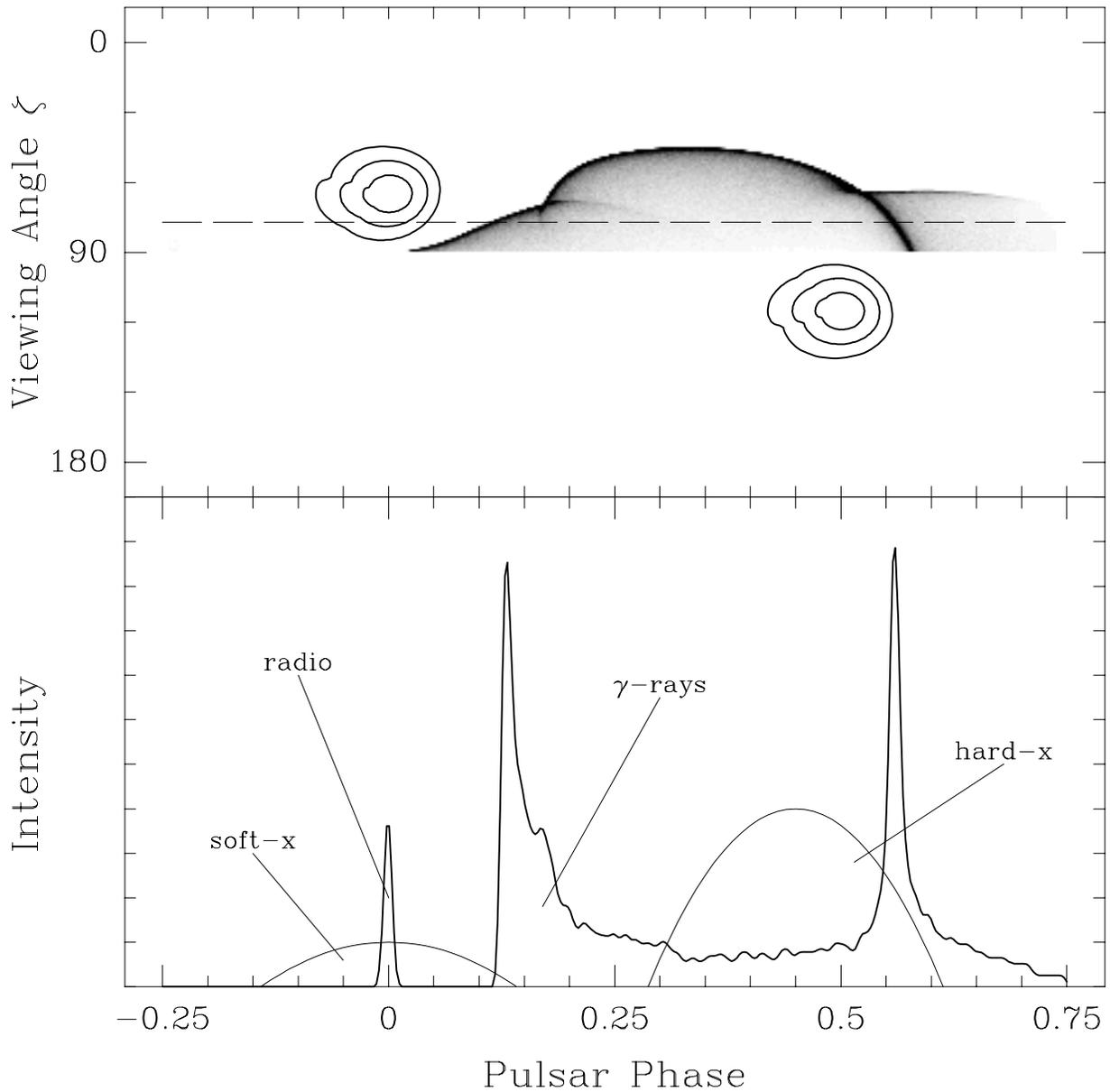}{6.0truein}{0}{85}{85}{-260}{-125}
\caption{
Skymap of pulsed beams and pulse profiles for Vela parameters
($\alpha=65^\circ$).
Above: Pulsar emission as a function of pulsar phase - observed
lines-of-sight cut horizontally across the image at constant $\zeta$; the
Earth line-of-sight is indicated (dashed line). Emission from the observed
outer gap (halftone) and both radio caps (contours) are shown. For clarity,
the outer gap visible for $\zeta > 90^\circ$ is not shown. Below: Pulse
profiles for Vela. The model $\gamma$-ray pulse for the line of sight above
is indicated, along with the leading radio pulse and schematic plots
of the hard (magnetospheric) and soft (thermal) X-ray pulses.}
\label{graymap}
\end{figure}

	With the emission zone defined, we note the radiating
particles will have significant $\gamma$ factors along the field lines.
High energy photons will thus be radiated in a narrow beam tangent to
the emission zone. As $E_\gamma$ decreases below an MeV, an increasing
fraction of the emission will arise from lower energy pairs and the
inverse-Compton and synchrotron photons producing the observed pulse
will have a larger spread about the field line (CR94). However, significant
flux will still be along the local field line and so the
high energy beam shape is defined by the tangent to this last closed surface.
Aberration effects and light travel delays will be very significant high
in the magnetosphere, so we account for these and bin the observed
radiation into pulse profiles. We display this as a skymap in pulsar phase and
observer co-latitude ($\zeta = \alpha + \beta$). In Figure~\ref{graymap}
we show the radio caps (from both hemispheres) along with the
$\gamma$-ray beam from the Southern radio cap. Angles and the indicated
line of sight are those appropriate to the Vela pulsar.

\subsection{Gamma Ray Beam Characteristics and Radio Data}

	Several features of the $\gamma$-ray beams are worth
mentioning. First, since radiation occurs above the null charge
surface, where ${\bf \Omega \cdot B} = 0$, $\gamma$-ray emission observed
at a given $\zeta$ comes from the polar cap in the opposite hemisphere.
Particles of the opposite sign from those producing the observed
emission will be accelerated inward in the gap. Their radiation would be
visible in the same hemisphere as the radio cap, however CR94 show that
the radiation efficiency and spectrum can be quite different for the
inward-going particle; we expect that they will produce little
observable flux.

	Second, as noted by Chiang and Romani (1992),
double pulses with significant bridge emission are visible for most
$\zeta$ as long as $\alpha$ is not too small. These pulses are caused
by crossing caustic surfaces in the pulse phase sky map, where flux from
a large portion of the outer gap arrives in phase. These caustics are produced
only with the full effect of aberration and light travel time and can
be quite sharp when the emission
surface is thin. Relatively narrow $\gamma$-ray pulses will be
visible for lines of sight close to the upper edge of the gamma ray beam,
and will have two poorly defined peaks with strong bridge emission, such as
PSR1706-44 and possibly PSR1055-52. Also very wide (width between $\gamma$-ray
peaks $\Delta > 0.5$)
double pulses are possible in this model, especially for small $\alpha$ and
\begin{figure}
\plotfiddle{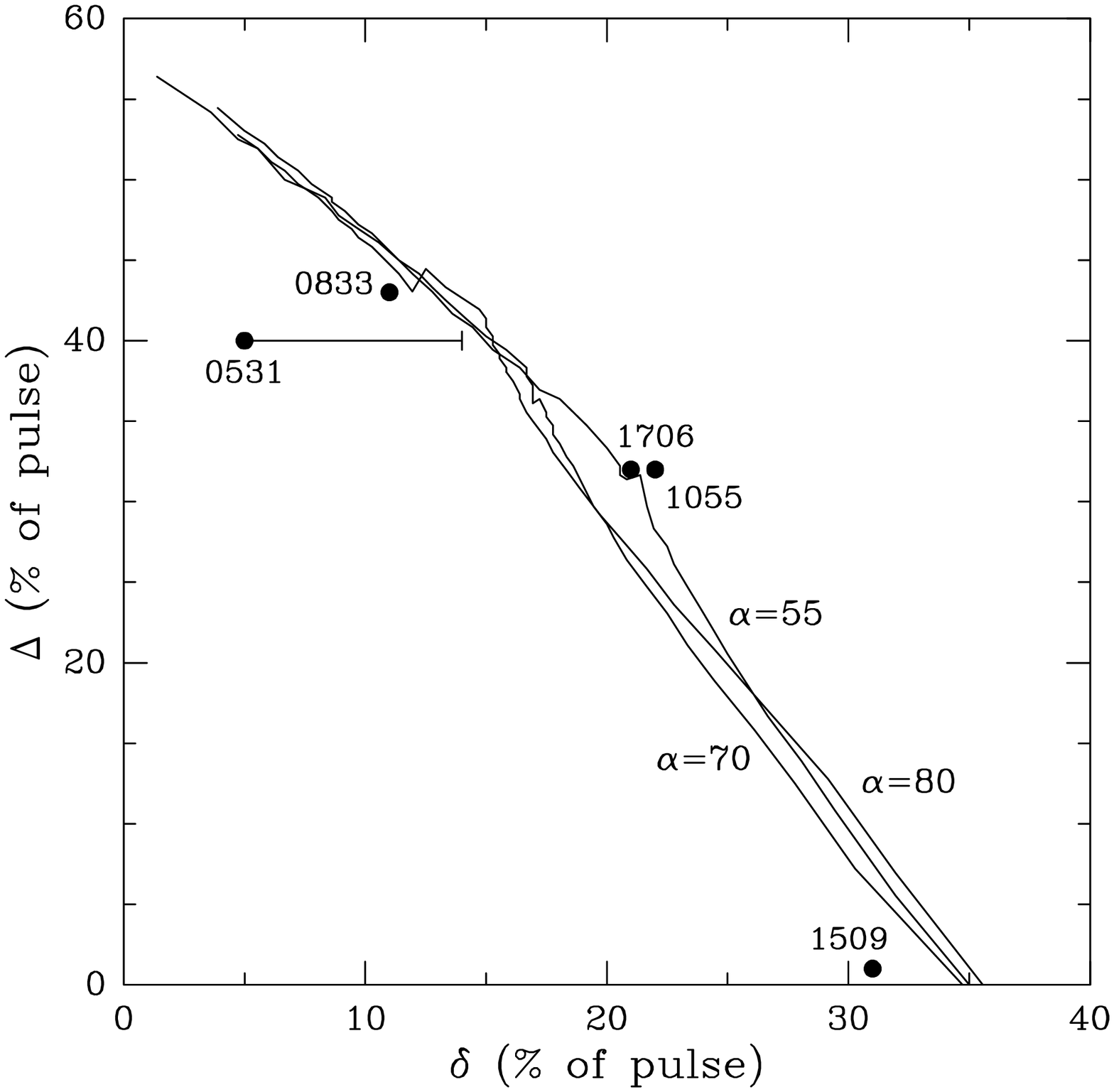}{4.5truein}{0}{60}{60}{-190}{-82}
\caption{
$\gamma$-ray pulse width $\Delta$ {\it vs.} lag of first $\gamma$ pulse
from the radio pulse $\delta$. Observed values for radio pulsars
are indicated. Model predictions for $\alpha$ covering the observed pulsar's
range are given by the full lines. The phase of the magnetic axis for the
Crab is somewhat uncertain -- the horizontal bar gives 1/2 the range over
which the core component is expected to appear.}
\label{dd}
\end{figure}
$\zeta \sim 90^\circ$. As $\alpha$ approaches $0^\circ$, regions where the
outer
gap approaches the light cylinder after crossing over the rotation pole
become increasingly important in the pulse profile. Faint emission from
this region is visible on the trailing side of the pulse in
Figure~\ref{graymap}.
Near $\alpha =0^\circ$, the $\gamma$ emission is concentrated to the
equator, and the region where strongly pulsed radiation is visible will be
small. Those pulsars seen at small $\alpha$ will tend to have wide
pulses.

	Radio data are particularly important in constraining possible
viewing geometries. In particular, polarization data can give very
good determinations of $\beta$ from the rotating vector model (Lyne and
Manchester 1988, LM).  Pulse widths can also be used to estimate $\alpha$
in some cases (Rankin 1990, 1993), but
in general $\alpha$ is only poorly constrained, unless emission is
visible in an interpulse or through a large range of pulsar phase.
However, the radio data (especially with accurate polarization measurements)
also give good measurements of the phase of the magnetic axis.
The relative phase of the radio and $\gamma$ pulse are now
available for the observed pulsars ({\it c.f.} Ulmer 1994).
In Figure~\ref{dd} we compare the $\gamma$-ray pulse width (between peaks) as
predicted from our model as a function of the phase offset from the radio
with the values measured for the observed $\gamma$-ray pulsars (Table 1).
The agreement is very good, although in the case of the Crab, the
radio emission lies slightly closer to the $\gamma$ pulse than expected.
However, we note that for this shortest period pulsar, the envelope in
which the radio emission, identified with the precursor, can appear is
relatively large. Half this range is shown by the horizontal bar.
For the other pulsars more complex profiles and polarization data allow
the true centroid of the radio pulse to be better determined.
Overall, the good reproduction of the observed correlation is a strong
success of the model.
\begin{figure}
\plotfiddle{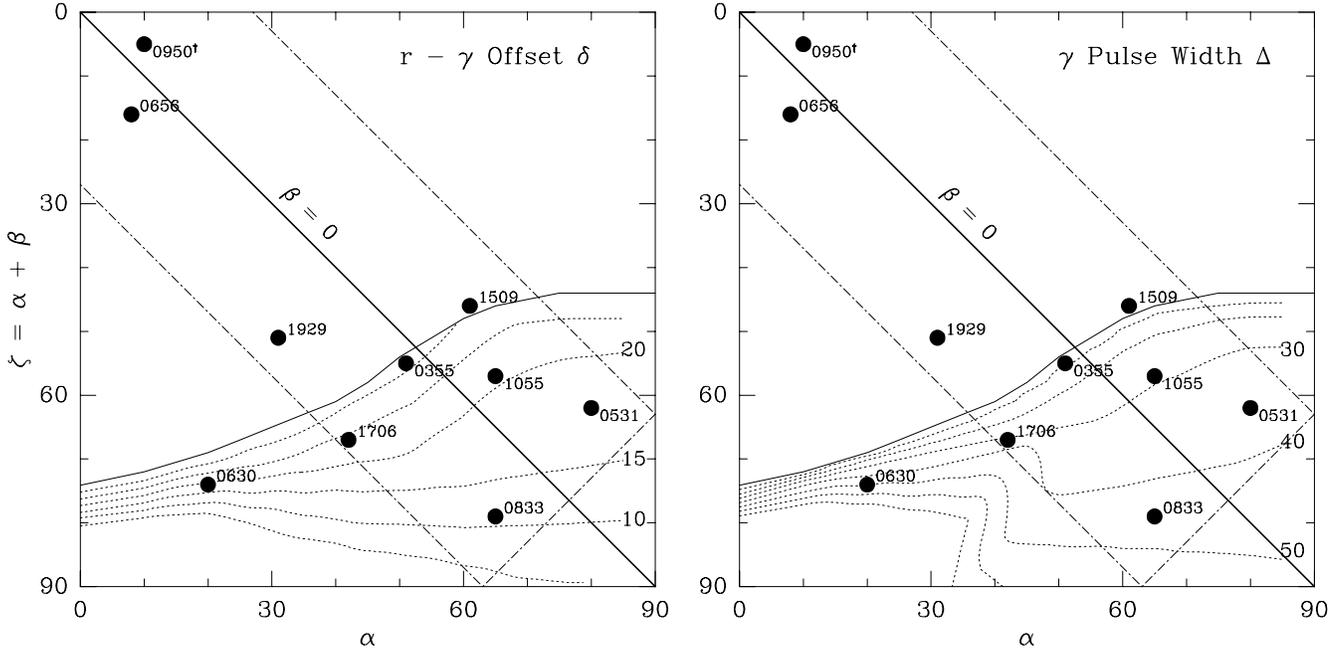}{3.4truein}{-90}{72}{72}{-305}{345}
\caption{
Isocontours of lag of from the radio phase $\delta$ (left) and
$\gamma$-ray pulse width $\Delta$ (right)
in the $\alpha - \zeta$ plane (see text). Contours are
computed for a gap width appropriate to a median age ($P \sim 0.1$s)
pulsar; selected contour values are labeled near $\alpha=90^\circ$.
Observed pulsars
are positioned at estimated values of ($\alpha,~ \zeta$); for Geminga
a large range along $\Delta=0.5$ is allowed. Pulsars known to
have $\zeta > 90^\circ$($\dagger$) have been displayed reflected to the
Northern hemisphere. Above the line $\Delta=0$ no $\gamma$ rays are visible;
far from $\beta=0$ pulsars will not be detected in the radio.}
\label{contplot}
\end{figure}

\subsection{Individual Pulsars }

	We have computed models for a range of $\alpha$ and sampled the
pulse profiles of these for the full range of $\zeta$. Figures~\ref{contplot}a
and \ref{contplot}b
show contours of constant $\delta$ and $\Delta$ respectively in the
$\alpha-\zeta$ plane. The gap width used in these sums is appropriate to
a $\sim 0.1$s pulsar, near the middle of the observed range.
Above the line of $\Delta=0$, no $\gamma$-ray emission is seen. Radio
emission is visible for small $\beta$; lines showing the range of
visibility for the Crab are drawn, although at low frequencies and/or low
intensities, radio flux may be visible over a wider range. Interpulsars,
with $\alpha$ near $90^\circ$ and radio emission visible from both poles
are in the lower right corner of the diagram. For $\alpha > 90^\circ$
the geometry will be similar with $\beta$ of oppposite sign.

	Using $\beta$ values (and $\alpha$, where available) from the
radio data we have plotted positions of young pulsars on Figure~\ref{contplot}a
and \ref{contplot}b.
In some cases, we have chosen a value of $\alpha$ consistent with
the radio and $\gamma$-ray data. For Geminga a substantial range along
$\Delta = 0.5$ is consistent with the data. Results
for each of the pulsars are in good agreement with the models. Moreover several
peculiarities of the observed profiles are explained with these figures.
For example, it is apparent that for much of the parameter space giving
$\Delta \ga 0.5$, as for Geminga, $\beta$ will be too large for the radio
pulse to be visible. Also for nearly aligned pulsars such as PSR1929+10
(Phillips 1990),
no $\gamma$ pulse is expected, so the strong EGRET upper limits can not
be construed to imply a luminosity cut-off at the characteristic age of
PSR1929+10.
Also, in several cases pulsars are located just outside the range of
$\gamma$ visibility ({\it eg.} PSR1509-58 and possibly
PSR0354+55); in these cases, the GeV $\gamma$-ray beam tightly collimated
to the gap surface may be missed, while the lower energy emission
with its wider beam pattern may be detected. In these cases OSSE, or
even X-ray energies, may provide the best hope of pulse detection.

	Most importantly, this model gives significant predictions for
the behavior of the radio polarization. Careful measurements should be
able to check these expected viewing geometries in many cases.
Conversely, polarization studies in connection with our high energy
model computations can help decide when careful pulse searches of
young pulsars are likely to be successful. Ultimately, knowledge of the
viewing geometry will help in locating the emission region for various
sectors of the pulse profile and, through phase-resolved spectra, will
allow a detailed study of the emission region.

\subsection{X-ray and Optical Pulses}

	In this paper we do not compute X-ray (0.1-10 KeV) profiles
in detail. However, we note that there are several important sources of
X-ray emission in this model. First we expect the region of the gap near
the null charge line to be a significant source of X-ray flux for young
pulsars. This flux will radiate widely and will be visible as a broad
`hard' pulse. We predict that such flux will appear at phases
$\sim 0.35-0.5$ after the radio pulse. \"{O}gelman (1994) has noted such
a hard `magnetospheric' pulse component in young pulsars and has shown
that it connects spectrally onto the high-energy power-law emission.
A second X-ray pulse will result from softer flux with a roughly
blackbody spectrum emitted from hotter regions near the magnetic axis on
the young star's surface, due both to anisotropic magnetic opacities
and to heating by
backflowing particles from acceleration zones (Arons 1983). These soft pulses
will lie near phase 0, but should again be quite broad, due to wide
emission zones and general relativistic bending of the photons near the
neutron star surface. The radio emission at $\sim 10-100$ stellar radii
should not suffer this effect. Several young pulsars, have in fact been
observed to have hard and soft X-ray pulses shifted in phase, in broad
agreement with this picture ($eg.$ PSR1055-52: \"{O}gelman and Finley 1993;
PSR0656+14: C\'ordova \et 1991, Finley \et 1992; Geminga: Halpern and
Ruderman 1993; and Vela: \"{O}gelman 1994).
When $\alpha$ is close to $90^\circ$ emission
may be visible from both caps and from both gaps, complicating the
profiles. A third, primarily unpulsed soft X-ray component can be
present due to the initial cooling of the hot neutron star; following
\"{O}gelman (1994) we expect this to be significant only for young
($10^4$y$\,\le \tau_c \le 10^6$y) neutron stars where the surface $T$ is
still large, but the gap luminosity is relatively small.

	At present optical pulse profiles are available only for the
Crab, Vela and PSR0540-69 pulsars. For the Crab, correspondence of the
optical and $\gamma$-ray profiles indicates that optical emission is
produced over much of the outer gap. However, for the Vela profile the
optical pulse is substantially narrower than that of the high energy
emission. The phase and width of the pulse are consistent with an origin
near the null charge surface ({\it c.f.} Figure~\ref{graymap}).
PSR0540-69 has a broad
pulse with two peaks (Middleditch and Pennypacker 1985, Gouiffes \et 1992)
consistent with a line of sight
cutting fairly high on an outer gap. Optical detections have however
also been established for Geminga (Halpern and Tytler 1988) and
suggested for PSR0656+14 and PSR1509-58 (Caraveo \et 1994a,b). For the
former two pulsars the optical emission may be from the
Rayleigh-Jeans tail of the thermal flux from the heated polar cap,
although non-thermal gap emission is preferred. PSR1509-58, however, if
correctly identified at $m_r \sim 21$ must be dominated by non-thermal
gap emission. This counterpart is bright enough to allow determination
of an accurate pulse profile; in our model we would expect a single
broad peak of emission aligned with the hard X-ray pulse ({\it i.e.} at
phase $\sim 0.3$ with respect to the radio peak).

\section{Crab Polarization Results}

	Interpretation of the Crab pulsed emission has long proved
problematic. It is clear from the alignment across a wide spectrum that
the radio-optical-X-$\gamma$ emission from the main pulse and interpulse
arises from similar zones, which we identify with the outer
magnetosphere gap (although the intensity weighting along the gap
surface will differ significantly for the different energy bands). We
follow other authors in identifying the precursor with the normal radio
pulse. Confusion arises however because the optical pulse exhibits a strong
double sweep in the polarization position angle -- this has long
been interpreted as surface emission in a two-pole RVM fit with
$\alpha \approx 90^\circ$ ({\it e.g.} Narayan and
Vivekanand 1982). Since the Crab is the only source for which polarization
information is presently available from the high energy region, this
association has been very influential in the interpretation of other
$\gamma$-ray pulsars.

	Adopting the same premise as the rotating vector model, (Radhakrishnan
and Cooke 1969) namely that the polarization vector of the emission is set
by the local magnetic field, we can
similarly compute the position angle swing through the pulse in our
outer gap picture. We compute the vector radius of curvature of the field line
at the emission point, project this on the plane of the sky, follow
aberration variations, and average the position angles from all
regions contributing to a given pulse phase to obtain the model
\begin{figure}
\plotfiddle{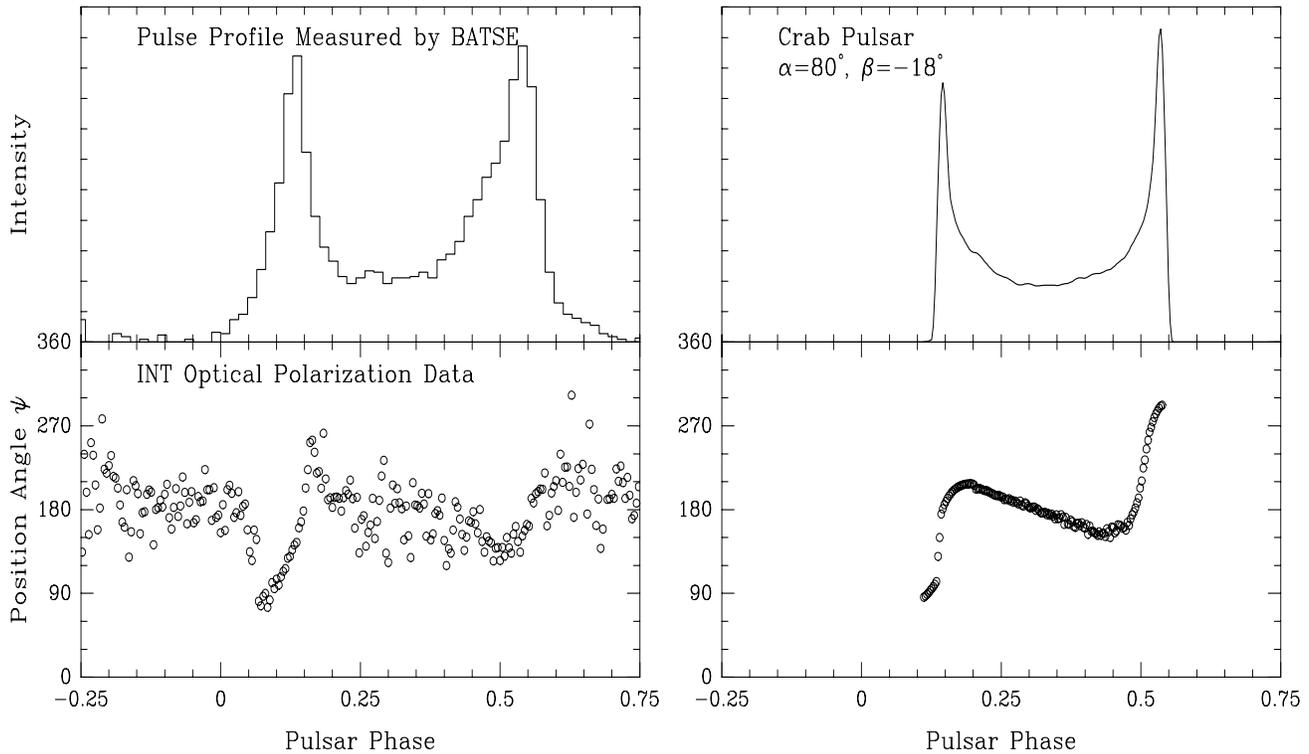}{3.9truein}{-90}{70}{70}{-293}{352}
\caption{ Crab pulsar profiles and polarization sweep. Left: light curve data
from BATSE (Fishman, GRO Newsl.,{\bf 1}, 6), INT optical polarization data
(Smith \et 1988).  Right: Model results for Crab outer gap.}
\label{both2}
\end{figure}
polarization sweep. Remarkably, because of the combination of emission
from a range of altitudes in the magnetosphere, our model produces an
initial sweep from the leading edge of the outer gap (the first pulse),
a reversal and a final sweep in the second pulse. This behavior is
obtained precisely for the viewing angles that produce the observed $\sim
140^\circ$ profile width. The left panels of figure~\ref{both2} shows
the optical polarization
position angle variation (from INT data, Smith, \et 1988) along with a high
energy pulse profile from BATSE.
Figure~\ref{both2} also shows model results with the
viewing angle chosen only to match the observed pulse width and
separation. While it is gratifying that the pulse profile matches well
to the data, it is remarkable that the corresponding polarization sweep
shows a striking similarity to the observed data, as well. We consider
this a major success of our model. With 10m class telescopes and modern
detectors it should be possible to obtain polarization data for Vela,
PSR1509-58 and PSR0540-69, as well. These would be particularly
important in motivating an emission location in the lower portions of the
outer gap -- since double sweeps generally occur only for $\alpha \ga 70^\circ,
{}~ \zeta \la 70^\circ$, we expect most of these pulsars to show a single
sweep of position angle.

\section{Population Estimates}

\begin{figure}
\plotfiddle{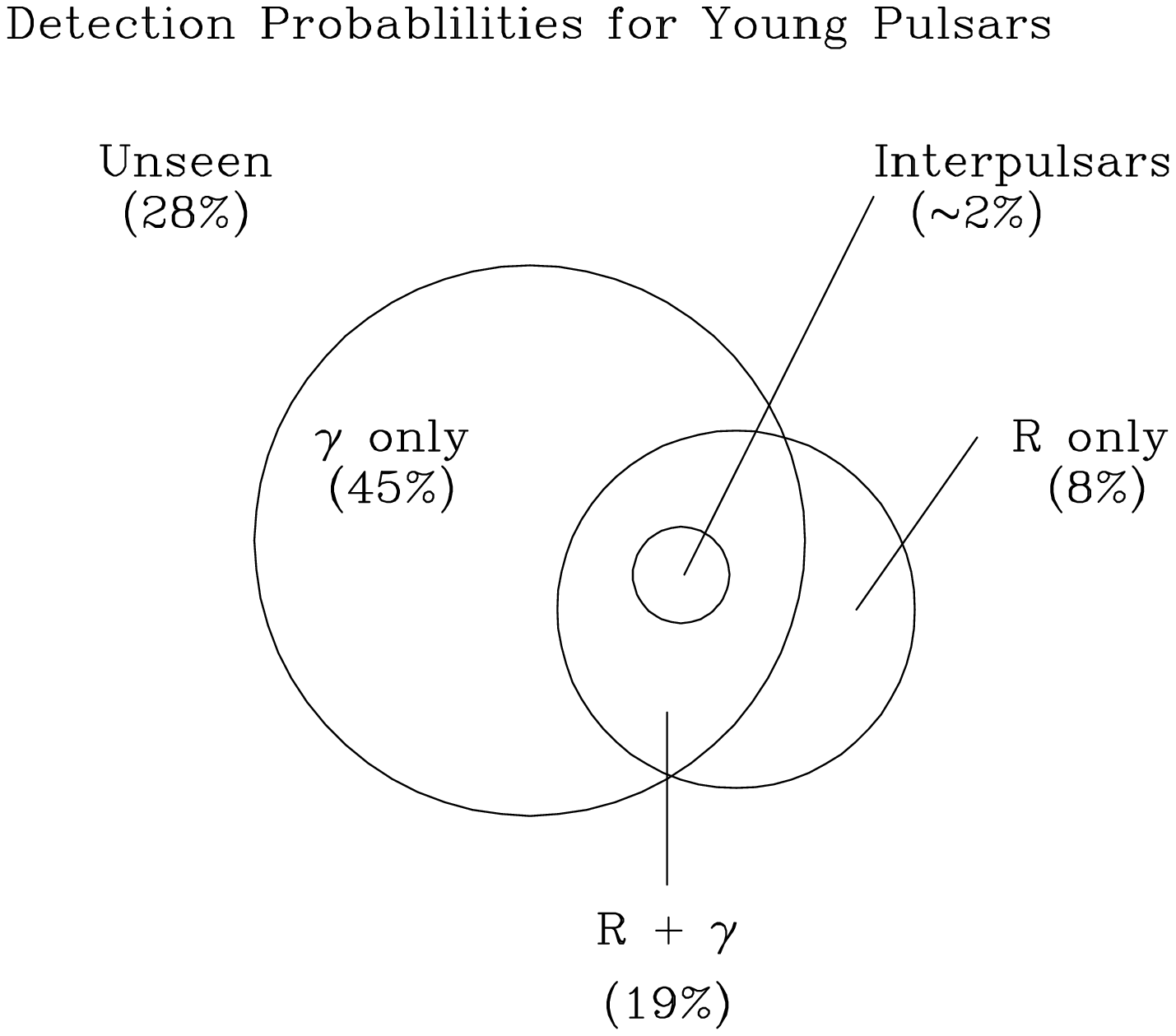}{3.7truein}{0}{70}{70}{-245}{-185}
\caption{
Venn diagram for pulsar detectability (beaming fractions). Random
magnetic inclinations, high energy emission from both poles
and radio beaming factors from LM are assumed. Values are from an integration
over pulsars with ${\rm Log}(B)=12.5$ and $\tau_c < 10^{5.5}$y.}
\label{venn}
\end{figure}

	In addition to allowing acceptable models for individual
pulsars, it is important for a $\gamma$-ray emission picture to
reproduce the population as a whole. To test our model on this front,
we have assumed that magnetic inclinations are random at birth
and that the observer orientation is also random and have integrated
the beaming fractions and detection probabilities for the radio
and $\gamma$-ray emission over a population of young ($\tau < 10^{5.5}$y)
pulsars. For the radio emission we have adopted conventional beaming
fraction relations ($W = 6.5^\circ P^{-1/3}$ LM;
$W = 5.8^\circ P^{-1/2 }$ Rankin 1993). The results for the fraction of
all young pulsars detectable in the two bands are shown in Figure~\ref{venn}
for
the LM beaming law; values for Rankin's beaming law differ by only
a few percent.
We have also considered how the results might be affected by alignment
torques secularly decreasing $\alpha$, but with typical alignment timescales
$\sim 10^7$y (LM) the differences in detection fractions are very small.
Uncertainty in the $P-{\dot P}$ dependence of the gap width makes these
fractions slightly uncertain; improved estimates of gap closure can remedy
this.

	Our population estimates show good agreement with present limited
data. First, as is well known, the mean radio beaming fraction of these
young pulsars, 0.27, leads to a pulsar birthrate reasonably consistent with
the galactic supernova rate. Second, roughly 1/3 of the radio selected pulsars
should not be visible in $\gamma$-rays, to a consistent luminosity threshold.
Of the young radio pulsars monitored, several do in fact have upper limits
on their efficiency of $\gamma$-ray production lower than that of the known
emitters (Thompson 1994). Thus with $\sim 5$ radio pulsars
detected in $\gamma$-rays it is not surprising that significant upper limits
exist for PSR0656+14 and PSR1929+10; non-detection for $\sim 2$ pulsars
should be due to $\gamma$ beaming. There is also a very strong upper
limit for PSR1951+32, but this is probably a mildly recycled pulsar
and has a dipole field roughly an order of magnitude lower than the young
pulsars. Because of decreased synchrotron emissivity and difficulty in
gap closure and inverse Compton scattering photon production, we would
not expect this to be an efficient $\gamma$-ray pulsar. True millisecond
pulsars are similarly poor targets for magnetospheric $\gamma$-ray emission.

	Finally, because
of the large $\gamma$-ray beams, the existence of objects such as Geminga
not detected in the radio is not surprising. Indeed, scaling from the
the 5 radio pulsar detections, we expect $\sim 12$ non-radio pulsars to
be visible in $\gamma$-rays to flux levels comparable to the faintest
radio selected object, in agreement with the observationally motivated
estimates of Helfand (1994). We thus expect most of the $\sim 20$ unidentified
galactic point sources to be young pulsars. In most cases identification
of the pulse period will be very difficult: X-ray emission, such as that used
to find the period of Geminga (Halpern and Holt 1992) offers the
best hope, but most
sources will be 3-10 times more distant, so the count rates will be low.
Support for the pulsar hypothesis may be obtained by finding the expected
hard Inverse Compton Scattering spectra above 100MeV, or in some cases by
association with old supernova remnants.

	It is important to note that theories placing the principal
$\gamma-$ray production in acceleration gaps at the polar cap
(Daugherty and Harding 1982, Harding, Ozernoy and Usov 1994;
Sturner and Dermer 1994) have great difficulty in meeting the
population test. These models cannot easily produce the range of
pulses with bridge emission observed. In particular, to produce wide pulses
these models require very small $\alpha \la 10^\circ$ with the observer line
of sight $\zeta$ nearly aligned with the rotation axis ${\bf \Omega}$
to intercept this small pole (Harding and Daugherty 1993,
Dermer and Sturner 1994). Even $if$
pulsars are preferentially born with aligned magnetic axes (a hypothesis
not supported by the radio data), the present significant number of detections
coupled with the small beaming fraction in these models
already implies an enormous birthrate of young pulsars, typically
$\ga 30 \times$ the galactic supernova rate.
We consider this to be unacceptable.

	Thus while some $\gamma$-rays may be produced in polar gaps,
this cannot account for the emission presently observed from radio pulsars.
The energetic arguments used in these models to estimate $\gamma$
fluxes will carry over to the outer gaps in many cases.
The processes converting the primary current to the observed $\gamma$
emission will however be quite different. Accurate models are not yet available
for the outer magnetosphere (CR94), so that luminosity laws for
$\gamma$-ray pulsars must at present remain phenomonological.

\section{Conclusions}

	We have computed the expected pulse shapes and beaming fractions
for $\gamma$-ray emission from young radio pulsars in a model
of charge acceleration in the outer magnetosphere. The results bear
a strong resemblance to $\gamma$-ray pulse profiles detected by CGRO.
In addition our model reproduces the phase relationships between the
radio emission and the $\gamma$-ray pulse. Observed X-ray and optical
pulsations find a natural interpretation in our geometric picture, as
well. Important connections between
radio pulse and polarization properties resulting from the
viewing angle and expected $\gamma$-ray profiles have also been established --
these give useful predictions for individual pulsars and allow strong tests
of the model. We further find that the polarization properties of the
Crab pulsar, with emission arising in the upper magnetosphere are easily
explained in our model; this supercedes previously confusing interpretations
of the polarization data. Finally, the numbers of objects detected in
the radio and $\gamma$-ray channels are as expected for a population of
young pulsars and we infer that most of the unidentified galactic plane
sources will be pulsars.

	These results are based principally on the {\it geometry} of the
emission region and are thus relatively free from uncertainties in details
of the emission process.  We are in the process of generating quantitative
models for the radiation produced in the upper magnetosphere, but
complicated non-linear models will be needed to give accurate results.
Previous estimates of outer gap spectra and fluxes (CHR88, Ho 1990) give
rough estimates of the total energy available, but do not give accurate
results for the observable flux or spectrum.  Nonetheless, the dominant
physical processes in this outer gap have already been identified by CHR88.
The substantial amendments needed for a realistic model, however, mean
that spectral results are not yet available.

	We feel that the success of our geometrical sums firmly establishes
the location of the $\gamma$-ray production in the outer magnetosphere
and thus resolves the long standing debate with polar cap models in
favor of an outer gap picture.  Unless similar results can be duplicated
by polar models, we feel that these models are not viable and efforts
to compute the spectrum and luminosity from the polar cap site
cannot explain the bulk of the observed pulsar emission.
Much work remains to be done to provide a full picture of the origin of
pulsar gamma rays, but assignment of the radiation site gives good
hope for further progress. In particular CR94 show that the radiation
processes vary strongly with altitude, and that these variations can be mapped
directly to pulse phase. Thus phase resolved spectra provide a keen
diagnostic for refining global emission models. Ultimately, detailed
comparison with observed profiles should allow us to probe the inertia
and current perturbations that are not followed in the present
calculation. Understanding
of these will help greatly in unraveling the mechanics of the
pulsar acceleration process and in producing a self-consistent model
of the pulsar magnetosphere. Thus $\gamma$-ray measurements can become an
important tool in deciphering the puzzle of the pulsar phenomenon.

\acknowledgments

	Support for this work was provided by NASA grants NAGW-2963
and NAG5-2037. We are grateful to Jim Chiang and Joe Fierro for important
discussions on the EGRET observations that improved this paper. We
also thank F. Graham Smith and Jamie Biggs for sharing the Crab optical
data.

\clearpage

\begin{table}
\begin{center}
\caption{Young Pulsar Parameters}
\vspace{1 cm}
\begin{tabular}{lrrrrrrrrrr}
PSR			&
$P$(ms)			&
$\log B(G)$		&
$\log \tau(y)$		&
$\log \dot{E}({\rm erg /s})$		&
$L_{\gamma}/\dot{E}$\tablenotemark{a}      &
$\delta$\tablenotemark{b}           &
$\Delta$\tablenotemark{c}               &
$\alpha(^\circ)$\tablenotemark{d}               &
$\beta(^\circ)$			&
X-rays\tablenotemark{e} \\
\tableline
\\
\multicolumn{11}{l}{$\gamma$-ray Pulsars} \\
\tableline
B0531$+$21 &  33 & 12.6 & 3.10 & 38.65 & 0.004 & 0.05 & 0.40 &  80 & -18 & h \\

B1509$-$58 & 150 & 13.2 & 3.19 & 37.25 &0.012$^f$&0.31 & $\sim$0& 60 & -15$^d$ & h \\
B0833$-$45 &  89 & 12.5 & 4.05 & 36.84 & 0.012 & 0.11 & 0.43 &  65& +14 & s,h \\
B1706$-$44 & 103 & 12.5 & 4.24 & 36.53 & 0.024 & 0.21 & 0.32 &  42 & +25 & s?,h \\
B0630$+$17 & 237 & 12.2 & 5.53 & 34.54 &$\sim$0.1-0.3& & 0.49 &  25$^d$ & +50$^d$ & s,h \\
B1055$-$52 & 197 & 12.0 & 5.73 & 34.48 & 0.2   & 0.22 & 0.32 &  70 &  -8 & s,h \\
\tableline
\\
\multicolumn{11}{l}{Candidates} \\
\tableline
B0540$-$69 &  50 & 12.7 & 3.22 & 38.17 &     &    & $\sim .3^g$    &     &     & h \\
B1951$+$32 &  40 & 11.7 & 5.03 & 36.57 &     &    &    &     &     & ... \\%
B0656$+$14 & 385 & 12.7 & 5.05 & 34.58 &     &    &    &  8  &  +8  & s? \\
B0355$+$54 & 156 & 11.9 & 5.75 & 34.66 &     &    &    &  51 &  +4  & ... \\%
B1929$+$10 & 227 & 11.7 & 6.49 & 33.59 &     &    &    &  31 & +20 & s? \\%
B0950$+$08 & 253 & 11.4 & 7.24 & 32.75 &     &    &    & 170 &  +5 & ... \\%
\tableline

\end{tabular}
\tablenotetext{a}{assumed beaming fraction 0.5 }
\tablenotetext{b}{radio pulse -- first $\gamma$ ray peak seperation (phase units)}
\tablenotetext{c}{$\gamma$ pulse width (phase units)}
\tablenotetext{d}{estimated from $\gamma$-ray data, when available. Values uncertain for Geminga.}
\tablenotetext{e}{s=soft flux from surface/heated polar cap, h=hard flux from base of outer gap}
\tablenotetext{f}{0.1-1 MeV flux only -- not detected by EGRET}
\tablenotetext{g}{estimated from X-ray, optical, radio data.}
\end{center}
\end{table}


\end{document}